\documentclass[aps, prd, twocolumn, superscriptaddress, preprintnumbers, nofootinbib, longbibliography]{revtex4-1}

\usepackage{amsmath}	
\usepackage{amsthm}		
\usepackage{amssymb}	
\usepackage{datetime}
\usepackage{graphicx}
\usepackage{color}
\usepackage{verbatim}
\usepackage{bm}

\usepackage[colorlinks=true, citecolor=midblue, linkcolor=midblue, urlcolor=midblue]{hyperref}

\usepackage{setspace}
\usepackage{ragged2e}

\usepackage[svgnames]{xcolor}

\usepackage{tikz,xcolor}
\definecolor{lime}{HTML}{A6CE39}
\DeclareRobustCommand{\orcidicon}{
	\begin{tikzpicture}
	\draw[lime, fill=lime] (0,0) 
	circle [radius=0.16] 
	node[white] {{\fontfamily{qag}\selectfont \tiny ID}};
	\draw[white, fill=white] (-0.0625,0.095) 
	circle [radius=0.007];
	\end{tikzpicture}
	\hspace{-2mm}
}
\foreach \x in {A, ..., Z}{
	\expandafter\xdef\csname orcid\x\endcsname{\noexpand\href{https://orcid.org/\csname orcidauthor\x\endcsname}{\noexpand\orcidicon}}
}

\settimeformat{ampmtime}

\definecolor{grey}{rgb}{0.4,0.4,0.4}
\definecolor{dullmagenta}{rgb}{0.4,0,0.4}
\definecolor{darkblue}{rgb}{0,0,0.4}
\definecolor{midblue}{rgb}{0,0,0.5}
\definecolor{midred}{rgb}{0.5,0,0}
\definecolor{orange}{rgb}{1,0.5,0}
\definecolor{lightbrown}{rgb}{0.75,0.5,0.25}
\definecolor{tan}{cmyk}{0.14,0.42,0.56,0}
\definecolor{djunglegreen}{cmyk}{0.99,0,0.52,0}
\definecolor{lightgreen}{rgb}{0,1,0}
\definecolor{olivegreen}{cmyk}{0.64,0,0.95,0.40}
\definecolor{midgreen}{rgb}{0.0,0.675,0.0}
\definecolor{darkgreen}{rgb}{0,0.5,0}

\newcommand{\Tr}{\ensuremath{\mathrm{Tr}}}

\settimeformat{ampmtime}

\usepackage{caption}
\usepackage{subcaption}

\usepackage[notrig]{physics}
\usepackage{tensor}

\newcommand{\FirstAffiliation}{\affiliation{
	Arnold Sommerfeld Center,
	Ludwig-Maximilians-Universit{\"a}t,
	Theresienstra{\ss}e 37,
	80333 M{\"u}nchen,
	Germany}}
 
\newcommand{\SecondAffiliation}{\affiliation{
	Max-Planck-Institut f{\"u}r Physik,
	Boltzmannstra{\ss}e 8,
	85748 Garching,
	Germany}}

 \newcommand{\ThirdAffiliation}{\affiliation{
        INFN, Sezione di Pisa, Largo Bruno Pontecorvo 3, I-56127 Pisa, Italy
	}}

\newcommand{\FifthAffiliation}{\affiliation{
Institut de F\'isica d’Altes Energies (IFAE) and The Barcelona Institute of Science and Technology (BIST),
Campus UAB, 08193 Bellaterra (Barcelona), Spain}}

\date{\formatdate{\day}{\month}{\year}, \currenttime}

\begin{document}

\title{Cosmological Implications of the Slingshot Effect:\\ Gravitational Waves, Primordial Black Holes and Dark Matter}

\author{Maximilian Bachmaier\orcidB{}}
\email{maximilian.bachmaier@physik.uni-muenchen.de}
\FirstAffiliation
\SecondAffiliation

\author{Gia Dvali}
\FirstAffiliation
\SecondAffiliation

\author{Juan Sebasti\'an Valbuena-Berm\'udez\orcidA{}}
\email{juanvalbuena@ifae.es}
\FifthAffiliation

\author{Michael Zantedeschi\orcidC{}}
\email{michael.zantedeschi@pi.infn.it}
\ThirdAffiliation

\date{\small\today} 

\begin{abstract}

In this paper, we explore the implications of the so-called 
\textit{slingshot} effect. It represents a rather general phenomenon occurring when a localized source, such as a monopole, quark, or a $D$-brane, crosses a domain wall separating the confined (Higgsed) and unconfined (Coulomb) phases of the crossing source. The crossover is accompanied by a stretched ``string''  of proper co-dimensionality that confines the source to the 
domain wall. The effect takes place for different setups, 
such as phase transitions leading to confinement, both electric and magnetic, as well as in string theoretic inflation with $D$-branes. We discuss the role of the phenomenon in sourcing gravitational waves and dark matter in the form of Kaluza-Klein gravitons. We also show that the slingshot effect can lead to the formation of primordial black holes in observationally interesting mass ranges for dark matter and high-energy cosmic rays.

\end{abstract}

\maketitle

\section{Introduction} 
This work is a continuation of the paper~\cite{Bachmaier:2023wzz}, where we study a phenomenon called \textit{slingshot}. It occurs in theories with coexisting confining and unconfining vacua. We analyze an unconfined charge that collides with the boundary of the confining vacuum. 

In a domain in which charges (electric or magnetic) get confined, the corresponding flux gets trapped in flux tubes. In the case of magnetic confinement, the flux tubes represent the Nielsen-Olesen vortex lines. On the other hand, in the case of the electric confinement, which takes place in $SU(N)$ gauge theories, the roles of the flux tubes are played by the QCD strings. 

\interfootnotelinepenalty=10000
When the two phases coexist and are separated by a domain wall, a charge placed on the confining side of the wall is attached to the wall via a flux tube. The tube opens up on the other side and spreads in the form of a Coulomb flux (electric or magnetic) into the unconfined domain. Simultaneously, the gauge field gets localized on the Coulomb side. This mechanism of localization of the massless 
gauge field was proposed in~\cite{Dvali:1996xe} and was further studied in~\cite{Dvali:2002fi, Dvali:2007nm}.
It was also noticed that the appearance of the localized massless gauge field, accompanied by the existence of the open strings 
attached to the wall, creates an obvious similarity with $D$-branes of string theory~\cite{Polchinski:1998rq}. And indeed, as discussed in the above papers, this connection has a deep physical meaning\footnote{The connection between $D$-branes and the $SU(N)$ domain walls of~\cite{Dvali:1996xe} is also evident from the mapping of their large-$N$ scaling properties on the QCD string theory~\cite{Witten:1997ep}. This scaling is supported by the exact large-$N$ solutions obtained in~\cite{Dvali:1999pk, Dvali:1999pk}.}.
In particular, in~\cite{Dvali:2002fi} various electric and magnetic examples of the above phenomenon were considered, and their connection with $D$-brane dynamics was discussed. In our study, we shall use a simplified prototype setup of this sort. 
\interfootnotelinepenalty=0
 
In~\cite{Bachmaier:2023wzz}, we focused on the case of a magnetic monopole in the Coulomb phase, scattering against a domain wall beyond which the vacuum is Higgsed and thereby monopoles are confined.  
Since the massless gauge boson carrying the field of the charge becomes massive, the flux cannot be immersed in the confined phase.
Consequently, when a monopole encounters this boundary, a magnetic flux in the form of a string emerges, connecting the monopole to the boundary. This flux tube serves as a channel for the residual field, effectively confining the magnetic charge to the wall between the two phases. After the collision with the wall, the charge slows down, and its kinetic energy is converted into the potential energy of the string.
Subsequently, the string reaches a maximal length permitted by the initial kinetic energy of the charge.
In the limit of a planar static wall, the string starts to shrink, pulling the charge back towards the wall. 
We refer to this phenomenon as the \textit{slingshot effect}, due to an obvious analogy with such an effect. 
While we focus on the case of magnetic monopoles, obviously, the slingshot effect can also occur for quarks, in scenarios where the color-confining and unconfining phases can coexist, as well as for 
$D$-branes~\cite{Dvali:1996xe, Dvali:2002fi, Dvali:2007nm}. 

Whether or not a long string is formed depends on the explicit realization of the scenario and the spectrum of the theory.  
In general, the string can break into shorter segments by nucleating pairs of opposite charges. In the case of a QCD string, this 
can be viewed as the process of  ``hadronization" during which the string decays into a multiplicity of mesons. On the other hand, in the limit of heavy charges, the pair creation is an exponentially suppressed tunneling process, and the strings become metastable. 
For a sufficiently large hierarchy, the break-up time is exponentially longer than the typical time scale of the Universe's evolution. In such cases, for all practical purposes, the string can be considered stable. We shall focus on such regimes.

In this work, we also investigate the interactions between two separated slingshots, with one monopole and one antimonopole entering the confining phase in parallel. Due to the long-range nature of the magnetic field in the Coulomb phase, we expect the string-wall junction points to attract each other, as we confirm with numerical simulations.

Our results have important cosmological implications due to the possibility of coexisting confined and unconfined phases in the early Universe. For QCD, this coexistence was discussed in~\cite{Witten:1984rs}. Moreover, such scenarios are generally expected to be realized in first-order phase transitions at intermediate symmetry-breaking scales of grand unified theories.

In~\cite{Preskill:1979zi, Zeldovich:1978wj}, it was found that the monopole production in the early Universe would lead to an overproduction problem. In these works, it was pointed out that monopole-antimonopole annihilation via Coulomb attraction is not efficient enough to reduce the monopole density to a level where it no longer dominates the energy budget of the Universe. This conclusion has recently been confirmed by numerical simulations~\cite{Hindmarsh:2025vxh}.
While a universal solution could be provided by inflation~\cite{Guth:1980zm}, in many motivated inflationary scenarios, the monopoles are formed after inflation. Correspondingly, the understanding of post-inflationary scenarios is very important. Such mechanisms that reduce the cosmological abundance of monopoles are provided by symmetry non-restoration~\cite{Dvali:1995cj}, and the erasure of magnetic monopoles via domain walls~\cite{Dvali:1997sa,Pogosian:1999zi,Brush:2015vda,Bachmaier:2023zmq,Senjanovic:2025enc}.

Another solution was suggested by Langacker and Pi~\cite{Langacker:1980kd}, who proposed that for a finite time interval, monopoles could be connected by strings, causing them to be pulled together and annihilate. As we show in this work, such string formation can also arise during first-order phase transitions through the slingshot mechanism, enriching the physical picture by Langacker and Pi.

All these ideas predict the production of gravitational radiation, allowing the study of the very early Universe through gravitational wave observatories. They can give information about whether and how magnetic monopoles appear in the Universe. Gravitational wave emission of magnetic monopoles connected by strings has already been analyzed~\cite{Martin:1996cp, Dvali:2022vwh, Buchmuller:2023aus}.
The gravitational wave emission of a single slingshot has been discussed in~\cite{Bachmaier:2023wzz}. 
In this work, we extend the analysis to a full system of multiple slingshots and give the first observational constraints on this source. We find that the signal typically lies in the high-frequency regime.

In this work, we also discuss implications of the slingshot mechanism for string cosmology. We shall argue that this phenomenon is rather generic in the string-theoretic inflationary scenarios, since they are universally based on time-dependent $D$-brane backgrounds~\cite{Dvali:1998pa, Dvali:1999tq, Dvali:2001fw}. During the evolution, the slingshot effect is realized by fundamental strings as well as by $D$-strings which connect $D$-branes of various dimensionalities. We show that the pull-over 
of $D$-branes by the string tension generically is expected to result in
detectable gravitational waves and the production of Kaluza-Klein gravitons. As originally proposed in~\cite{Arkani-Hamed:1998sfv}, the Kaluza-Klein gravitons of a large extra dimension~\cite{Arkani-Hamed:1998jmv} can potentially account for today's dark matter.

Moreover, the formation of primordial black holes (PBHs)~\cite{Zeldovich:1967lct,Hawking:1971ei,Carr:1974nx,Chapline:1975ojl,Carr:1975qj} by the mechanism proposed in~\cite{Dvali:2021byy} is expected. The PBH masses obtained by this mechanism within $D$-brane cosmology are expected to be in the sub-asteroid mass range. Such PBHs, stabilized by the memory burden effect~\cite{Dvali:2018xpy, Dvali:2020wft, Alexandre:2024nuo, Dvali:2024hsb,Dvali:2025ktz}, can be viable dark matter candidates and simultaneously represent interesting sources for both gravitational wave as well as high-energy cosmic ray observatories~\cite{Fermi-LAT:2014ryh,LHAASO:2023gne,IceCube:2018fhm,IceCube:2020wum,KM3NeT:2025npi}. 

The rest of this work is organized as follows: the next section recapitulates the case of a single slingshot amply discussed in~\cite{Bachmaier:2023wzz}. Section~\ref{sec:two-slingshots} addresses the case of two slingshots and the relative interaction between the two junctions. We further discuss the impact of the relative twist between the monopoles in the dynamics. 
Section~\ref{sec:many-slingshots} is dedicated to the case of a full phase transition. Section~\ref{sec:gravitational-radiation} describes the gravitational wave signal from monopole slingshots in the early Universe, while Section~\ref{sec:implication-of-slingshot-in-string-cosmology} addresses the implication of the slingshot in string cosmology of $D$-brane inflation. Finally, Section~\ref{sec:discussion-and-outlook} is left for our conclusion and outlook. 

\section{One slingshot}
\label{sec:one-slingshot}
In order to present the slingshot effect in an example, we will take a gauge theory characterized by the breaking pattern 
\begin{align}
    SU(2)\rightarrow U(1) \rightarrow 1\,.
\end{align}
This two-step symmetry breaking requires two distinct scalar fields, an $SU(2)$ adjoint scalar field, $\phi$, and an $SU(2)$ fundamental scalar field, $\psi$.
The coexisting phases with confined and unconfined 
electric or magnetic charges in such a system 
were discussed in~\cite{Dvali:2002fi, Dvali:2007nm}. 
For our analysis, we adopt a prototype model of this sort.  
The potential of the theory is given by
\begin{align}\label{eq:Potential}
    U(\phi,\psi)&=\lambda_\phi \left(\Tr (\phi^\dagger \phi)-\frac{v_\phi^2}{2}\right)^2\nonumber\\
    &+\lambda_\psi \left(\psi^\dagger \psi- v_\psi^2\right)^2\ \psi^\dagger \psi+\beta \psi^\dagger \phi \psi\, ,
\end{align}
where we are assuming $v_\phi>v_\psi$.
A first symmetry breaking is done by the field $\phi$, allowing for the 't Hooft-Polyakov magnetic monopole solution~\cite{tHooft:1974kcl, Polyakov:1974ek}. 
The second breaking by the scalar $\psi$ leads to the existence of Nielsen-Olesen magnetic flux tubes~\cite{Nielsen:1973cs}. 

In the following, unless otherwise stated, we will work in units of gauge coupling $g=1$ and the gauge boson mass $m_{v_\phi}=g v_\phi=1$ coming from the symmetry breaking by the field $\phi$.

For a slingshot effect to occur, it is crucial that both the confined and unconfined phases can coexist. In equation~\eqref{eq:Potential}, this is realized by the sextic form of the $\psi$ potential. The vacuum ${\abs{\psi}}=0$ describes the unconfined Coulomb phase, and $\abs{\psi}=v_\psi$ signifies the confined Higgsed phase. 

Notice that, while the potential is non-renormalizable, the sextic behavior can arise from a renormalizable theory 
after integrating out a heavy field (see, for instance, in~\cite{Dvali:2002fi}) and can also emerge 
naturally from thermal and quantum effects. It can be considered as an effective potential that describes an intermediate stage of a first-order phase transition near the critical point~\cite{Dvali:2022rgx}.

We verified the presence of the slingshot effect in this model through a numerical simulation. For that, we introduced a planar domain wall separating the unconfined and the confined vacua. For $\beta \neq 0$, there is a pressure difference between these two vacua, leading to the acceleration of the domain wall in the direction of the energetically lower vacuum. We chose $\beta$ such that it accelerates in the direction of the unconfined Coulomb vacuum.
The domain wall solution, which separates the Coulomb and the Higgs vacua, can be obtained by solving the Bogomolny equation~\cite{Bogomolny:1975de}.

In~\cite{Bachmaier:2023wzz}, we implemented a single magnetic monopole in the unconfined region. Since the massless gauge field in the Coulomb phase receives a mass in the Higgsed phase, the magnetic field described by $B_i=-1/2 \varepsilon_{ijk} \hat{\phi}^a G^a_{ij}$, with $G^a_{ij}$ being the field strength tensor, should exhibit a repulsive behavior away from the domain wall similarly to the Meissner effect in superconductors. 

For a single monopole, we realized this by using the monopole-antimonopole ansatz with a maximal twist~\cite{Saurabh:2017ryg}. The magnetic field resembles the right shape between the two monopoles. We excluded one of the two monopoles in our setup.

The time evolution was computed numerically using the Crank-Nicolson method described in~\cite{Teukolsky:1999rm}. 
In the axially symmetric cases, such as the configuration described above, we used a special technique that significantly reduces the computation time. With axial symmetry, it is sufficient to solve the field equations in one plane only and use the symmetry to obtain the values for the neighboring planes of the lattice. Here we were able to use a lattice spacing of $0.25 m_{v_\phi}^{-1}$ and smaller.
In non-axially symmetric scenarios, for example, as for a monopole and antimonopole entering the Higgsed region at different points (see Section~\ref{sec:two-slingshots}), we used a cubic lattice of length $240 m_{v_\phi}^{-1}$ with lattice spacing of $1.0  m_{v_\phi}^{-1}$. Because of the low resolution, after the slingshots formed, we turned on an artificial friction term in the gauge field equation for stability reasons.
For more details and explanations on the numerical implementation, we refer to the paper~\cite{Bachmaier:2023wzz}.

\begin{figure*}
    \includegraphics[trim=32 0 70 5,clip,width=\textwidth]{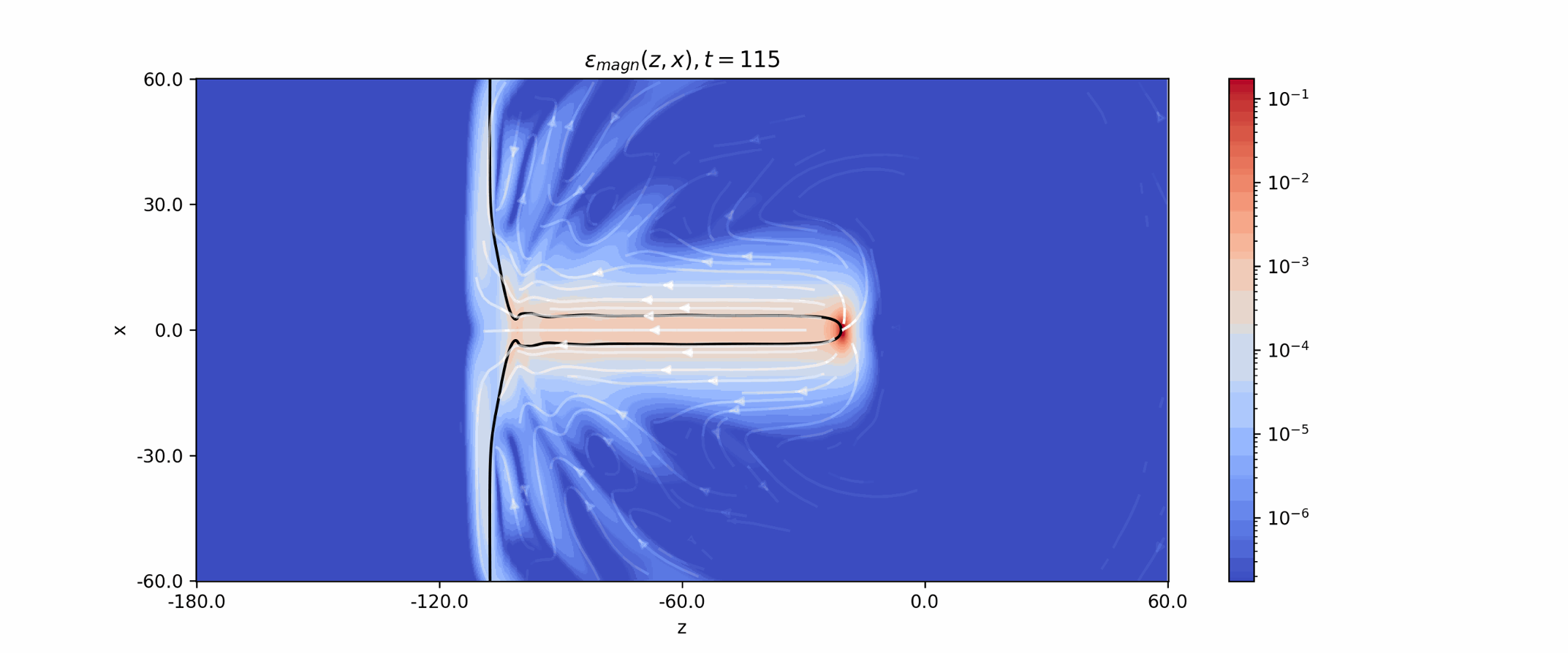}
    \caption{\justifying
    The magnetic energy density in the $y=0$ plane at time $t=115\, m^{-1}_{v_\phi}$ in units $m_{v_\phi}^4/g^2$. 
    The black lines illustrate the contour $\abs{\psi}=0.5\, v_\psi$. Here, we chose the following parameters: $m_{h_\phi}=\sqrt{2\lambda_\phi} v_\phi=m_{v_\phi}$, $m_{v_\psi}=g v_\psi /\sqrt{2}=0.15\, m_{v_\phi}$, $m_{h_\psi}=2\sqrt{\lambda_\psi}v_\psi^2=0.6\, m_{v_\phi}$, $\beta=0.01\, m_{v_\phi}$. This figure appeared first in~\cite{Bachmaier:2023wzz}.}
    \label{fig:magnetic-energy-density}
\end{figure*}

We ran the simulations for different parameters of order one. 
Because of the rapid acceleration of the domain wall, the collision happened in the relativistic regime. After the collision of the wall with a single monopole, the monopole immerses into the confined region and stretches a string that is connected to the domain wall. Fig.~\ref{fig:magnetic-energy-density} shows one time frame of the simulation.
The full-time evolution of this process and also the scenarios discussed in the next sections can be found in the following video:\\
\url{https://youtu.be/PnErf4-zUEg}

We can observe that once the string reaches its maximal length, $l_{\text{max}}=\gamma_c M/\mu_{\text{string}}$, where $\gamma_c$ is the relative Lorentz factor, $M$ is the monopole mass and $\mu_{\text{string}}$ is the string tension, the monopole moves with the domain wall at approximately the same velocity.

A natural question that can be asked is whether the string can detach from the domain wall. Thermal or quantum fluctuations, in fact, can lead to a breaking of the string, via the nucleation of a monopole-antimonopole pair~\cite{Vilenkin:1982hm}. Such a process is, however, exponentially suppressed for monopoles heavier than the confinement scale. 

Classically, the only way to achieve the detachment of a single string from the wall is to let an antimonopole enter it from the unconfined domain. We checked this in another numerical simulation. We observed that the antimonopole disconnects the string from the wall and forms a monopole dumbbell (a monopole-antimonopole pair connected by a string).
Subsequently, these monopoles undergo acceleration toward each other, leading to a collision and annihilation, as previously described in~\cite{Dvali:2022vwh}. 
It was observed that instead of recreation after the monopole-antimonopole collision, the pair completely decayed into the waves of Higgs and gauge bosons. Entropy arguments elucidated the reason for this behavior~\cite{Dvali:2020wqi}. The system prefers the state of highest entropy, favoring wave-like patterns over the monopole-antimonopole pair.
It is worth mentioning that in a cosmological setup, the annihilation can take place at extremely high center-of-mass energies, potentially leading to the formation of PBHs, as discussed in~\cite{Dvali:2021byy,Matsuda:2005ey}. 
This introduces an additional channel for PBH formation arising from topological defects. The production of PBHs from pure domain walls~\cite{Berezin:1982ur, Garriga:2015fdk, Deng:2016vzb, Ferrer:2018uiu, Gelmini:2022nim, Dunsky:2024zdo, Gouttenoire:2025ofv} 
and from pure cosmic string loops~\cite{Hawking:1987bn} has already been studied previously.

\section{Two Slingshots}
\label{sec:two-slingshots}
To understand how the slingshot effect appears in multi-monopole scenarios, it is important to examine the interaction between different slingshots. 
For a single slingshot, the string opens and releases the magnetic flux into the unconfined region at the string-wall junction. 
At the opening point, the magnetic field resembles that of a magnetic point charge.

We consider now the slingshot of a pair, by placing two monopoles separated 
along the $x$-axis and let them collide with an accelerating domain wall such that the two monopoles stretch two separated strings.

In contrast to a single monopole entering the confined phase, the magnetic field lines of the initial configuration for a monopole-antimonopole pair were already approximately oriented correctly. To refine these configurations for the simulation, we applied a numerical relaxation procedure to the initial setup\footnote{A description of how such a relaxation procedure can be applied to monopoles with fixed position can be found in~\cite{Bachmaier:2025jaz}.}.

We analyzed several different scenarios. For a monopole-antimonopole pair, we have also checked different twists, $\gamma$, which describe the relative phase between the monopole and antimonopole~\cite{Saurabh:2017ryg, Dvali:2022vwh}. Specifically, we simulated cases with twist $\gamma=0$, $\gamma=\pi/2$, and maximal twist $\gamma=\pi$.
We also studied a monopole-antimonopole pair where the two monopoles have different initial distances to the domain wall. 

\begin{figure*}
    \begin{subfigure}{\textwidth}
        \includegraphics[trim=20 0 80 10,clip,width=\textwidth]{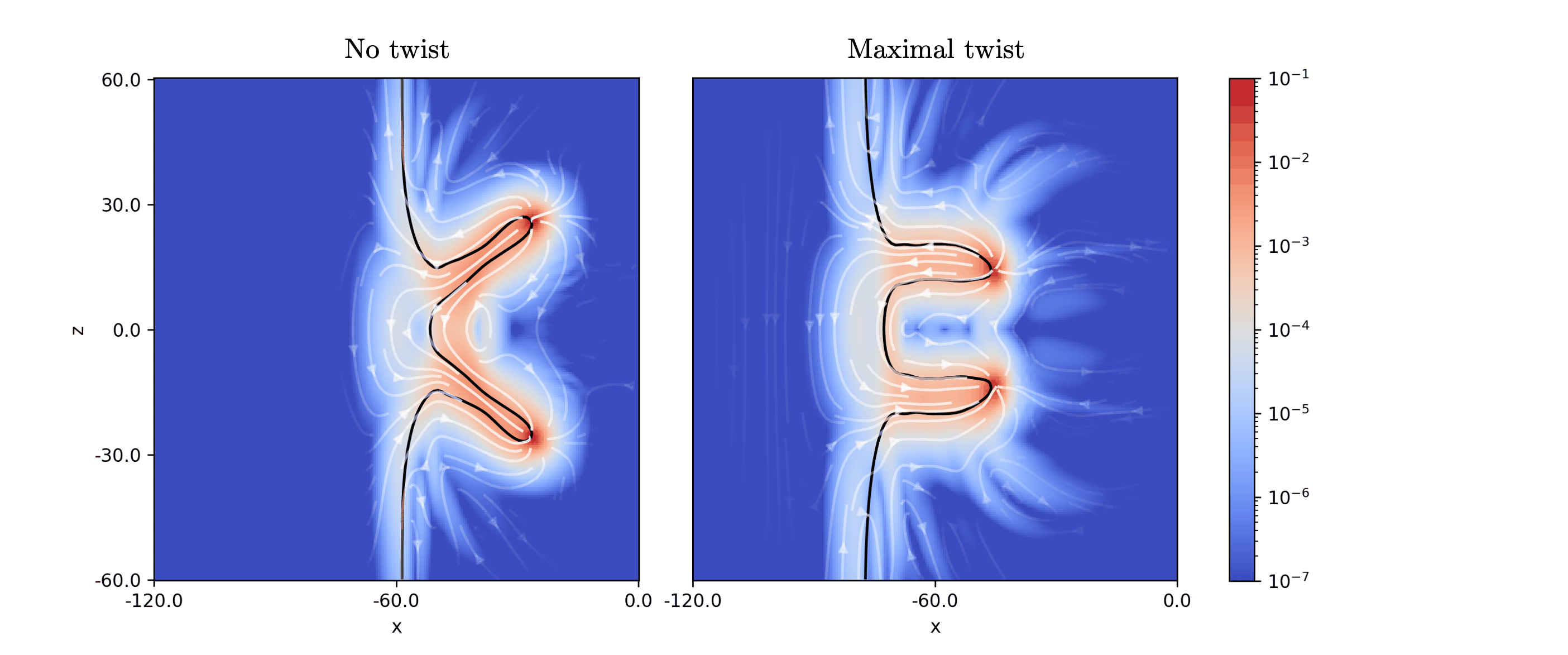}
    \end{subfigure}
    \begin{subfigure}{\textwidth}
        \includegraphics[trim=0 0 0 0,clip,width=\textwidth]{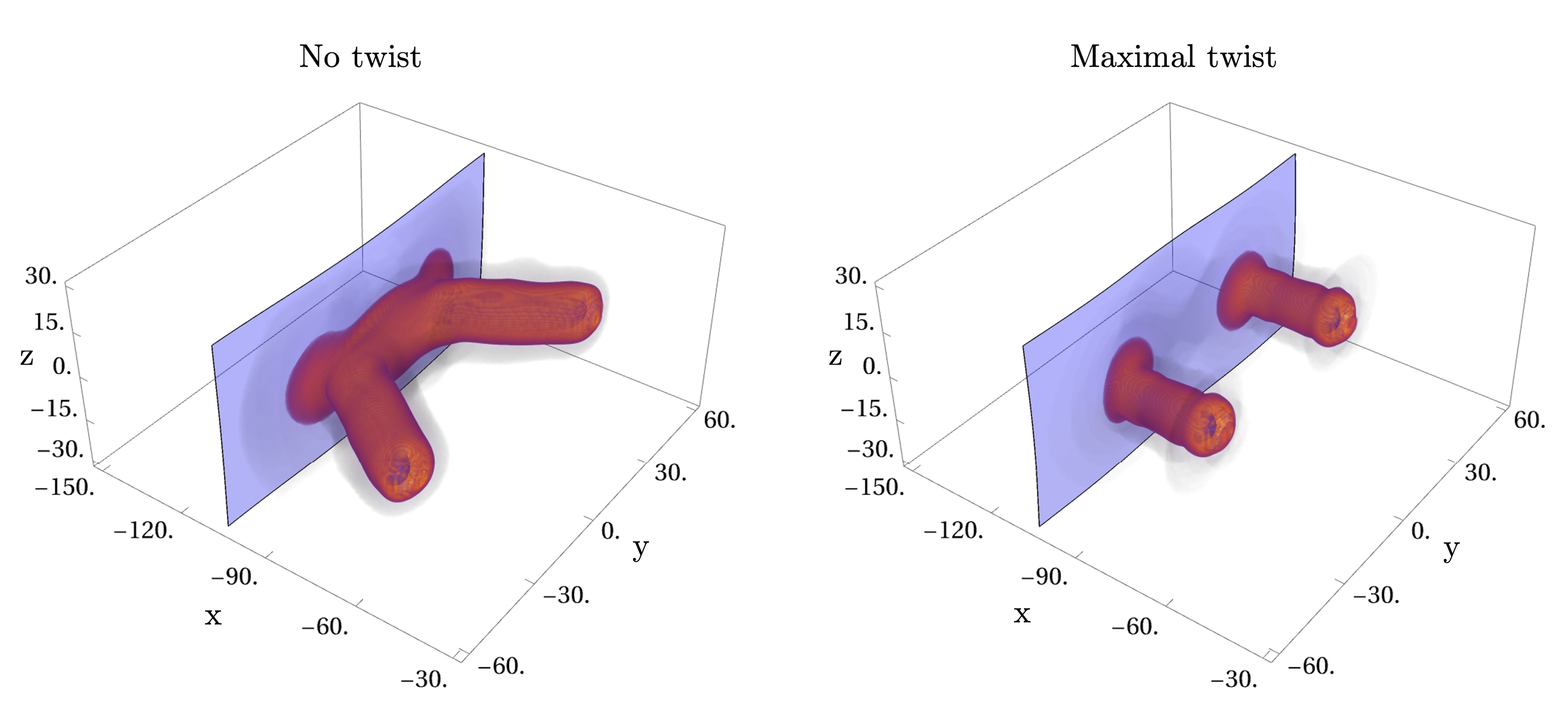}
    \end{subfigure}

    \caption{\justifying
    The upper two plots show one time frame of the magnetic energy density in the $y=0$ plane for the monopole and antimonopole with no relative twist (left) and with maximal twist (right) entering the Higgsed region.
    The black lines illustrate the contour $\abs{\psi}=0.16\, m_{v_\phi}$. The initial separation distance between the two magnetic monopoles was $d=34\, m_{v_\phi}^{-1}$ for both cases.
    The lower two plots show one time frame of two slingshots with initial separation $d=60\, m_{v_\phi}^{-1}$ with no relative twist (left) and with maximal twist (right). The red density plot shows the magnetic energy for values bigger than $5.0 \cdot 10^{-6}\, m_{v_\phi}^4/g^2$(left) and $2.0 \cdot 10^{-5}\,m_{v_\phi}^4/g^2$(right). The blue contour plot shows the domain wall profile value $\abs{\psi}=0.2\, m_{v_{\phi}}$.
    Here, we chose the following parameters: $m_{h_\phi}=m_{v_\phi}$, $m_{v_\psi}=0.2\, m_{v_\phi}$, $m_{h_\psi}=0.6\, m_{v_\phi}$, $\beta=0.01\, m_{v_\phi}$.}
    \label{fig:doubleslingshot}
\end{figure*}

In the confined region, the magnetic flux is concentrated within the string, but at the opening, the flux can enter the Coulomb phase in which the magnetic interaction is long-range. Hence, for a monopole-antimonopole pair, the opening points should attract, whereas for two monopoles with the same charge, they should repel.

In the simulations, we varied the initial separation between the monopoles from $d=20m_{v_\phi}^{-1}$ to $d=90m_{v_\phi}^{-1}$.\\

\noindent \textbf{No twist:} In the absence of twist, in the regime that we investigated, we observed that the magnetic flux from the string does not spread radially into the Coulomb region. Instead, the flux remains concentrated on the wall, forming a flux tube along its surface (see Fig.~\ref{fig:doubleslingshot}). This behavior was already mentioned for confined quarks in~\cite{Baldes:2020kam}.
As the distance between the monopoles increases, the flux tube on the wall becomes wider. For $d \gg 1/\Lambda$, where $\Lambda~\sim v_\psi$ is the confinement scale, the configuration approaches that of two isolated slingshots, where the magnetic field radiates outward from each string endpoint, resembling point-like magnetic charges on the wall.

Since the magnetic flux tube connecting the two monopoles tends to straighten, the strings of the slingshot bend accordingly, as illustrated in Fig.~\ref{fig:doubleslingshot} (left). This bending becomes less relevant at larger separation distances $d$ due to the broader distribution of the magnetic flux along the wall.

Once the string-wall junctions come together close enough, the strings detach from the wall and form a monopole-antimonopole pair connected by a string that attracts the monopoles until they collide and annihilate.\\

\noindent \textbf{Maximal twist:} For a maximal relative twist between the monopole and the antimonopole, the dynamics change significantly. Instead of forming a flux tube along the wall, the magnetic field extends farther into the Coulomb region. As a result, the strings remain straight and parallel throughout the evolution, as shown in Fig.~\ref{fig:doubleslingshot} (right). Once the strings approach each other closely, the string and antistring annihilate, leaving behind a twisted monopole-antimonopole pair connected by a very short string. Notice that in the maximally twisted scenario, it is not possible to form long strings from two parallel slingshots. This is because the detachment of the strings happens at a moment when the monopole and the antimonopole are already very close to each other.

After the formation of the twisted monopole-antimonopole connected by the string, the monopoles bounce once before they untwist and annihilate. 
This bouncing of the monopoles was already simulated in~\cite{Dvali:2022vwh}. In a symmetric setup, the monopoles may never annihilate and form a static sphaleron configuration~\cite{Klinkhamer:1984di}.

In the numerical simulations, we also examined non-maximal twist angles. In these cases, the magnetic field outside of the strings is partially concentrated in a flux tube and is partially spread into the Coulomb region. As a result, the field configuration is a superposition of the non-twisted and maximally twisted cases. Consequently, the bending angle varies smoothly with the twist parameter $\gamma$, from maximal bending at zero twist to no bending at maximal twist.

In all cases we analyzed, the collision was always perpendicular to the domain wall. This scenario was particularly interesting because after a phase transition in which magnetic monopoles emerge, they typically move slowly due to interactions with the surrounding plasma. In first-order phase transitions, however, domain walls expand rapidly, forming large bubbles within a short time. As a result, perpendicular collisions dominate in such phase transitions.

Our results of the simulations apply to scenarios in which the distance between the slingshots, $d$, is of the order of the string thickness. However, for large distances, $d\gg 1/\Lambda$, the attraction between the junctions of the slingshots becomes Coulomb-like. For $d\ll L$, with $L$ being the string length, the mass that is accelerated is just given by a small piece of the string close to the domain wall, $m\sim \mu_\text{string}/\Lambda\sim\Lambda$. For large distances, $d \gg L$, the full slingshot is moving, and so the accelerated mass is given by $m\sim \mu_\text{string} L\sim \Lambda^2 L$.
The motion is described by the equation $m \Ddot{x}\sim1/x^2$ whose solution gives the attraction time $\tau$ between two slingshots\footnote{The dynamics between the junctions might be altered due to the interaction with the plasma. In fact, this may lead to partial localization of the magnetic flux along the wall or 2-brane, leading to a logarithmic long-distance interaction (rather than Coulomb-like), thereby affecting the maximum number of monopoles that can simultaneously undergo the slingshot. Such a scenario is highly model-dependent and beyond the scope of this work.}
\begin{align}
\label{eq:slingshot-decay-time}
   \tau \sim \begin{cases}
       \Lambda^{\frac{1}{2}}\, d^{\frac{3}{2}}&\text{for}\, \,  d \ll L\,,\\
        \Lambda\, L^{\frac{1}{2}}\,d^{\frac{3}{2}}&\text{for}\, \,  d\gg L\,.
   \end{cases}
\end{align}

\section{Many Slingshots}
\label{sec:many-slingshots}
The previously constructed scenarios are well-suited to explore the details of the interaction between two slingshots. To get a better intuition of what happens during a full first-order phase transition in the Universe, we have also simulated a growing bubble in a system containing many monopoles created dynamically by a phase transition $SU(2)\rightarrow U(1)$. 

For that, we started with $\phi=0$ in the simulation. We added small perturbations on the scalar field $\phi$ such that it falls into the vacuum $\abs{\phi}=v_\phi$ dynamically. In this phase transition, monopoles and antimonopoles are formed.

During this process, much of the potential energy is converted into thermal (kinetic) energy. We added an artificial friction term to the field equations to damp these fluctuations.

As initial conditions, we input a small vacuum bubble in the $\psi$ field with radius $r=3m_{v_\phi}^{-1}$. We keep the $\psi$ field fixed while evolving $\phi$ and the gauge fields to produce the monopoles. After their production, we start evolving also the $\psi$ field, leading to the expansion of the bubble.

\begin{figure*}
    \includegraphics[trim=50 240 70 0,clip,width=\textwidth]{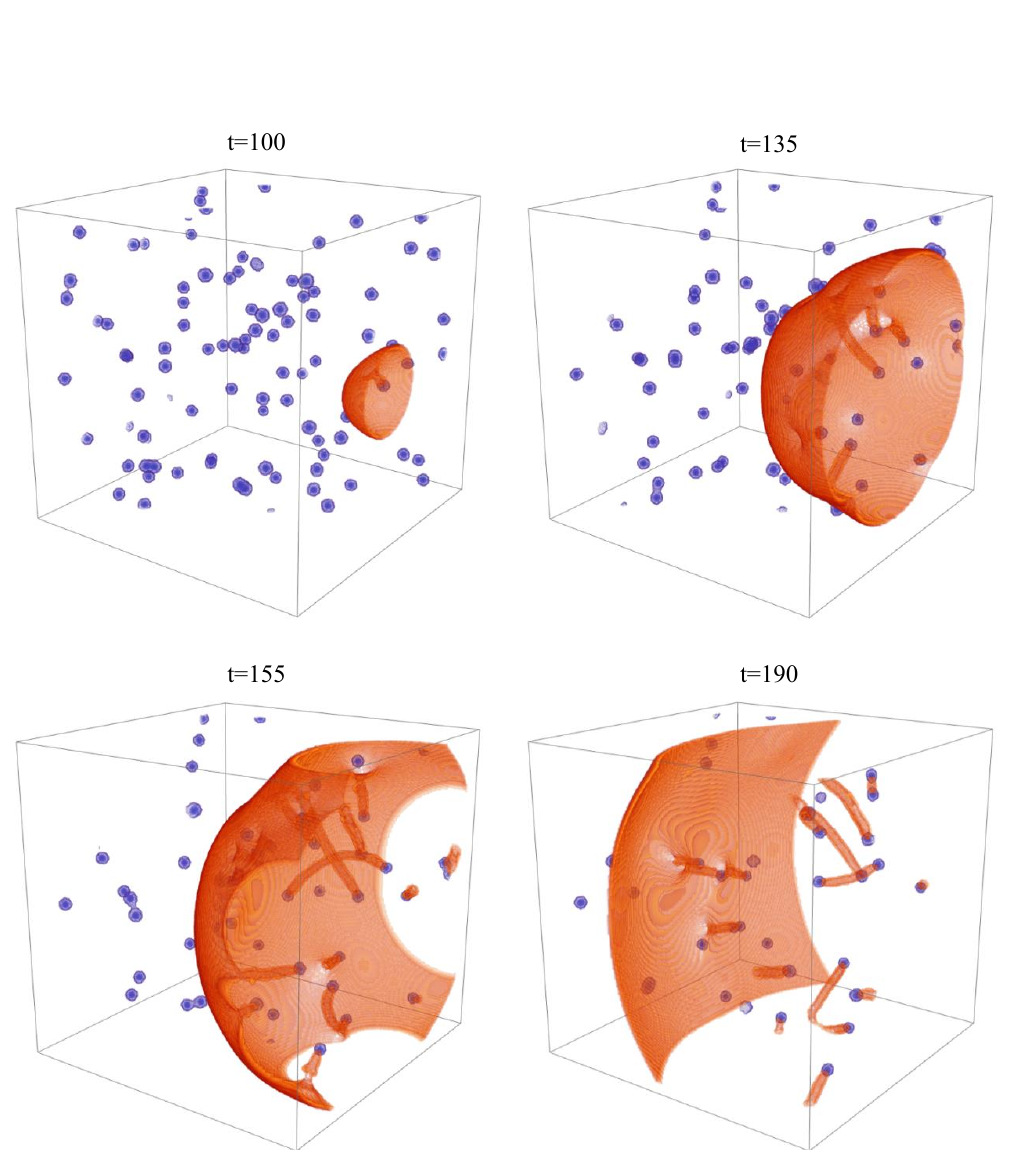}
    \begin{flushleft}
    \caption{\justifying The time evolution of a growing bubble that is colliding with magnetic monopoles, leading to many slingshot events. The blue density plot shows $\abs{\phi}$ and illustrates the position of the monopoles. The orange density plot corresponds to $\abs{\psi}$ and displays the domain walls and strings.
    The figure shows a box of size $(170\, m_{v_\phi}^{-1})^3$.
    The full simulation was performed in a lattice of size $180\, m_{v_\phi}^{-1}\cross 180\, m_{v_\phi}^{-1} \cross 360\, m_{v_\phi}^{-1}$ and lattice spacing $1.0\, m_{v_\phi}^{-1}$.
    Here: 
    $m_{h_\phi}=m_{v_\phi}$, $m_{v_\psi}=0.15\, m_{v_\phi}$, $m_{h_\psi}=0.6\, m_{v_\phi}$, $\beta=0.2\, m_{v_\phi}$.}
    \label{fig:phasetransition}
    \end{flushleft}
\end{figure*}

This procedure allows us to study what happens to magnetic monopoles when they collide with a vacuum bubble during a first-order phase transition. Thanks to the periodic boundary conditions, we could also observe the slingshot behavior during bubble collisions. In Fig.~\ref{fig:phasetransition}, we present an example of the simulation results. 

As we have already seen for two slingshots, we again observed the bending of the flux tubes and the attraction of the junctions. In addition, several important new phenomena emerged.

For example, we observed the formation of strings with both ends connected to the wall without monopoles involved.
Such configurations can form when two slingshots intersect. Due to the string tension, instead of forming a string loop in the confined phase, the string will eventually collide with the wall and effectively will be erased~\cite{Dvali:2022rgx}.

Another observed phenomenon occurs when two bubbles collide. 
Since monopoles always need to be connected by a string in the confined phase, very long strings form that connect slingshots originating from different bubbles. If the monopoles are heavy compared to the string, and the configuration does not intersect with other strings, this long monopole dumbbell is very long-lived. However, if the monopoles are light, the string tends to break into smaller segments through monopole-antimonopole pair creation via quantum tunneling~\cite{Vilenkin:1982hm}.\\

\section{Gravitational Radiation}
\label{sec:gravitational-radiation}
Due to the high accelerations that occur during the interactions between the magnetic monopole, the domain wall, and the string, a significant magnitude of gravitational radiation can be emitted. 
For a single slingshot, we computed the resulting gravitational radiation spectrum and found that it decays with $\sim \omega^{-1}$~\cite{Bachmaier:2023wzz} and that the peak of the emitted power scales as $P_{\rm GW} \simeq \Lambda^4 / M_{\rm P}^2$, where $\Lambda$ is the monopole confinement scale and $M_{\rm P}$ denotes the Planck mass. A similar result has already been obtained for a monopole-antimonopole pair connected by a string in the point-like approximation~\cite{Martin:1996cp} and in a fully field-theoretic simulation~\cite{Dvali:2022vwh,Bachmaier2026thesis}. 

We denote the temperature at which the slingshot takes place by $T_{\rm sl}$. 
The gravitational wave energy density generated by the slingshots can be estimated by
\begin{align}
\label{eq:omega-GW-ansatz}
\Omega_{\rm GW}^{\rm peak} \simeq P_{\rm GW}\, t\, n_{\rm M} \left( \frac{T_0}{T_{\rm sl}} \right)^4 \frac{1}{\rho_{\rm c}}\, ,
\end{align}
where $t$ is the duration of the slingshot, $n_{\rm M}$ denotes the number density of monopoles at the time of the events, $\rho_c$ is the critical density of the Universe, and $T_0$ is the CMB temperature.
Notice that we are assuming that the Universe is radiation-dominated until monopole annihilation, at $T\sim T_{\rm sl}$. We are further assuming no slingshot intersection/annihilation throughout the process (see comments below).
We will assume that the slingshot duration is given by the bubble collapse time. Since domain walls move with relativistic speeds, the duration is of the order of the horizon size.

Taking all this together gives the gravitational wave density parameter
\begin{align}
    \label{eq:OmegaGW}
    \Omega_{\rm GW}^{\rm peak}\simeq \frac{\Lambda^4\, n_{\rm M}}{T_{\rm sl}^6\, M_{\rm P}}\, \Omega_{\rm rad} \,,
\end{align}
where $\Omega_{\rm rad}$ is the radiation density parameter.
The frequency of today's signal generated by the slingshot is given by
\begin{align}
\label{eq:GW-frequency-signal}
    f\simeq \frac{1}{t}\frac{T_0}{T_{\rm sl}}\simeq \frac{T_{\rm sl}T_0}{M_{\rm P}}\,.
\end{align}

There are several possible scenarios involving the slingshot effect. The minimal one is the usual first-order phase transition in which bubbles nucleate, grow, and collide at $T \sim T_{\rm sl} \sim \Lambda$. 
Through supercooling, such a phase transition can be delayed, which means that the temperature at which the phase transition takes place can be $T \lesssim \Lambda$. Supercooling in confinement phase transitions has recently been discussed in~\cite{Agrawal:2025wvf}.
Another possibility is that domain walls form at $T \sim \Lambda$ and evolve into a scaling regime, of energy density $\rho_{\rm DW}\simeq \Lambda^3/t$ (see e.g., the review \cite{Saikawa:2017hiv}), for some time before collapsing or dissolving at $T \sim T_{\rm sl} < \Lambda$.
Because of these different possibilities, we work with $T_{\rm sl}$ and $\Lambda$ as separate parameters.

Several caveats need to be considered.
To avoid the so-called domain wall problem~\cite{Zeldovich:1974uw}, we require that the collapse of the walls happens at temperatures
\begin{align}
    T_{\rm sl}\gtrsim \left(\frac{\Lambda}{M_{\rm P}}\right)^{\frac{3}{2}}M_{\rm P}\, .
\end{align}
Furthermore, the monopole energetic domination is avoided for
\begin{align}
\label{eq:monopole-domination}
    n_{\rm M}\lesssim \frac{T_{\rm sl}^4}{M}\,.
\end{align}

In the derivation of equation~\eqref{eq:OmegaGW}, we assumed that the 
duration of the slingshot is given by the horizon time. However, due to the Coulomb interaction of the junctions discussed in Section~\ref{sec:two-slingshots}, this assumption holds only if $t\lesssim \tau$, where $\tau$ is the attraction time between two junctions, given in equation~\eqref{eq:slingshot-decay-time}. For $d\ll L$,  
the monopole density is bounded as
\begin{align}
    \label{eq:condition-collapse-faster-than-Coulomb-attraction}
        n_{\rm M}\lesssim \frac{T_{\rm sl}^\frac{14}{3}\, \Lambda^\frac{2}{3}}{M_{\rm P}^\frac{7}{3}}\, ,
\end{align}
where we used that the average distance between the junctions is given by $\langle d\rangle \sim 1/\sqrt{n_{\rm M}t}$.

Taking the Coulomb interaction into account, we can find that a first-order phase transition at $T_{\rm sl}\sim \Lambda$ can provide $\Omega_{\rm GW}^{\rm peak}\lesssim (f/T_0)^{10/3}\, \Omega_{\rm rad} $. A scenario in which the domain walls form at $T\sim\Lambda$ and evolve in a scaling regime until $T_{\rm sl}$ will give $\Omega_{\rm GW}^{\rm peak}\lesssim (f/T_0)^{16/9}\, \Omega_{\rm rad}$, where we used equation~\eqref{eq:condition-collapse-faster-than-Coulomb-attraction}. These results show that, assuming that all the above-mentioned bounds hold, the gravitational wave signals from the slingshot mechanism will not be observable by the future generation of gravitational wave experiments.
More non-minimal setups may strengthen the signal further, but a more detailed analysis is left for future work.
Obviously, in the case $t \gtrsim \tau$ (or in a case with non-zero initial slingshot velocity), the duration, length, and shape of the slingshot need to be properly accounted for in the above estimates. 
For example, inequality~\eqref{eq:condition-collapse-faster-than-Coulomb-attraction} can be weakened if the slingshot junctions interact with the plasma. Taking monopole energetic domination as a constraint gives 
$
    \Omega_{\rm GW}^{\rm peak}\lesssim \left({f}/{T_0}\right)^{2/3} \left({M_{\rm P}}/{M}\right)\,\Omega_{\rm rad}\, .
$
As an example, we can consider the monopole mass to be of the order of the grand unification scale, $M \simeq 10^{16}\, \rm GeV$. In this case, a signal with frequency $f \simeq 10^2\, \rm Hz$ (corresponding to $\Lambda \simeq10^{13}\, \rm GeV$) can yield a gravitational wave density parameter as large as $\Omega_{\rm GW}^{\rm peak} \simeq 10^{-8}$. Such a signal lies within the sensitivity reach of current detectors such as LIGO, Virgo, and KAGRA (LVK)~\cite{aLIGO:2020wna,VIRGO:2014yos,KAGRA:2018plz}, as well as future gravitational wave observatories like the Einstein Telescope~\cite{Sathyaprakash:2012jk} and Cosmic Explorer~\cite{LIGOScientific:2016wof}.

Overall, we can conclude that the gravitational-wave signal produced by the slingshot effect lies in the high-frequency regime, a range for which active efforts are made to develop new detection concepts~\cite{Aggarwal:2020olq,Franciolini:2022htd,Gatti:2024mde}.
In addition to the slingshot itself, there are, of course, several additional sources for gravitational radiation from a first-order phase transition involving magnetic monopoles. 
As previously mentioned, the strings detach from the wall and form monopole dumbbells. In the non-twisted case, the bending of the string remains after disconnection. Due to string tension, the string tends to straighten, causing it to accelerate and oscillate between the monopole and antimonopole, which results in the emission of gravitational radiation. Additionally, the head-on acceleration of the monopoles toward each other also produces gravitational waves. 

In our slingshot simulations, we have observed that monopoles connected to the wall by strings pull the wall inwards, leading to significant deformations of the wall. Such deformations also influence the emission of gravitational waves during collisions of vacuum bubbles~\cite{Megevand:2021juo}. Moreover, the monopoles act as a source of friction on the wall, reducing the acceleration of the bubble. Thus, the gravitational radiation from domain wall collisions and networks is modified from previous studies~\cite{Hogan:1986dsh, Caprini:2018mtu}.

\section{Implications of the Slingshot effect in $D$-brane Inflation}
\label{sec:implication-of-slingshot-in-string-cosmology}

The slingshot effect can have implications in $D$-brane cosmology of string theory, and in particular, it is expected to be rather generic in $D$-brane inflation~\cite{Dvali:1998pa, Dvali:1999tq, Dvali:2001fw}.
The basis of this scenario is cemented by the fact that, within the current understanding of string theory, $D$-branes are the only objects with negative pressure. Therefore, the inflationary epoch within string theory was likely dominated by $D$-brane dynamics.
 
The realization of $D$-brane inflation is based on some general model-independent features~\cite{Dvali:1998pa, Dvali:1999tq, Dvali:2001fw}. In particular, during inflation, the Universe has to be dominated by a configuration of $D$-branes that are significantly displaced from the current vacuum. This displacement creates the potential energy that dominates the Universe and leads to inflation. During inflation, the $D$-branes are slowly relaxing towards the vacuum. The degree of freedom describing the relaxation plays the role of the slowly rolling inflaton field. The simplest realization of this scenario is provided by several $D$-branes and anti-$D$-branes that are displaced relative to each other in extra dimensions. At each moment, the inflaton field corresponds to a combination of fields that account for distances between the branes in extra dimensions. 

Before moving to an actual discussion of a $D$-brane slingshot, let us briefly formulate a general framework. We shall assume that the relevant cosmological background is obtained by compactification of a $10$-dimensional string theory to four space-time dimensions. That is, we assume that six out of nine space dimensions are compact. $D$-branes differ by the dimensionalities of their world-volumes which have $p$ space dimensions, with $p < 10$.  The $10$-dimensional tension (the energy per unit world-volume) goes as ${\mathcal T}_p \sim g_{\rm s}^{-1}M_{\rm s}^{p+1}$, where $M_{\rm s}$ is the string scale and $g_{\rm s}$ is the dimensionless string coupling. Upon compactification, $D$-branes can be wrapped around different dimensions. For example, a $D_{3+n}$-brane wrapped around $n$ compact dimensions ($n$-torus) of radii $R_{j},~j=1,2,...,n$, from the point of view of a $3$-space-dimensional observer, will have an effective tension
\begin{equation} \label{Teff}
   {\mathcal T}_{3+n} \sim g_{\rm s}^{-1}M_{\rm s}^4(2\pi M_{\rm s})^n\prod_{j=1}^n R_j\,. 
\end{equation}  
By wrapping $D$-branes around different dimensions, one can create objects of different codimensions. 
 
It is conventionally assumed that during this epoch, the volume of extra dimensions is already stabilized, and the main motion takes place in a fixed extra space. This imposes 
two important conditions outlined in~\cite{Dvali:1998pa}. 
First, at any given moment of time, the effective four-dimensional Hubble radius, $R_{H}$, must not be shorter than the size of any extra dimension. 
That is, we must work in the regime, 
\begin{equation} \label{RHcondition}
R_{\rm H} \gtrsim R_j \,.
\end{equation} 
Secondly, the curvature of the stabilizing effective potential (the effective mass) of any radius modulus must not be smaller than the effective four-dimensional Hubble parameter 
\begin{equation} \label{RMcondition}
  m_R \gtrsim  H \equiv \frac{\Lambda_{\rm eff}}{3M_{\rm P}^2} \,,
\end{equation}
where $\Lambda_{\rm eff}$ is the effective four-dimensional energy density. The above conditions guarantee that on the relevant scales the evolution of the Universe can be reliably described by four-dimensional Friedmann-Lemaître-Robertson-Walker (FLRW) cosmology with the Hubble parameter $H$. In what follows, we shall assume that both of the above conditions are satisfied. 

We must note that in the current state of the art in string theory, the stabilization of extra dimensions as well as of other moduli represents an open problem, into which we shall not enter. 
This concerns both the present and the inflationary epochs. In this sense, brane inflation is not creating a new conceptual issue but rather extends the existing one to the inflationary epoch. 

In order to understand the slingshot dynamics, we must separate the longitudinal dimensions that include our three space coordinates, $x_{\parallel}$, and the transverse ones, $x_{\perp}$. 
Of course, $x_{\perp}$ are all compactified, whereas from $x_{\parallel}$, only the ones 
that are additional to our three space coordinates are compact.

With this convention, the world-volume dynamics of moving $D$-branes responsible for inflation is described in $x_{\parallel}$, whereas their relative motion takes place in some subset of extra coordinates $x_{\perp}$. Obviously, the world-volume space dimensionality 
of an inflating $D_{3+n}$-brane satisfies $n \geq 0$, with integer $n$.
Notice that for $n \geq 1$, this inflation-driving $D$-brane must at the same time be wrapped around $n$ of the extra compact dimensions belonging to $x_{\parallel}$.

In scenarios with large extra dimensions~\cite{Arkani-Hamed:1998sfv, Antoniadis:1998ig, Arkani-Hamed:1998jmv}, the compactification radius can be as large as $\sim 30\mu \rm m$, in accordance with the current bounds on Newtonian gravity~\cite{Adelberger:2009zz, Tan:2016vwu} as well as other constraints~\cite{Arkani-Hamed:1998jmv}.  
 
During their motion, the $D$-branes are going to stretch the strings that either connect them or end on the $D$-branes of lower dimensionality, which, from the point of view of a $3+1$-dimensional observer, appear as point-like objects. The surviving stretched strings can play the role of superheavy dark matter~\cite{Dvali:1999tq}. The schematic representations of the $D$-brane slingshot effect are shown in Fig.~\ref{fig:magnetic-energy-density} and Fig.~\ref{fig:doubleslingshot}, where the domain wall and the monopoles must be identified with the $D$-branes of respective dimensionalities.  
  
For simplicity, in the following, let us consider a single extra dimension $x_{\perp}$ compactified on a circle of radius $R_{\perp}$. We assume that the other five dimensions are compactified at much smaller scales, $R_j \sim M_s^{-1}$, and are irrelevant for the inflationary dynamics. We simultaneously take the number of wrapping dimensions $n=0$. The generalization to the $D$-branes of other codimensions is straightforward.  

In the implementation of the slingshot effect in $D$-brane cosmology, we can distinguish the two basis cases, to which we can refer as \textit{transverse} versus \textit{longitudinal} slingshot scenarios.

The longitudinal scenario concerns the slingshot effect that has significant projection in $x_{\parallel}$ coordinates. This slingshot will be driven by walls, strings, and monopoles as discussed in the previous section, with the only difference that these objects will have a string-theoretic origin. In other words, they shall be described by branes and fundamental strings with projected $x_{\parallel}$ codimensionalities 
of one, two, and three, respectively. Various objects that project in our dimensions $x_{\parallel}$ as strings have been discussed in~\cite{Dvali:2003zj, Copeland:2003bj}. In~\cite{Sarangi:2002yt}, it has been shown that they can be generically formed at the end of the brane inflation~\cite{Dvali:1998pa, Dvali:1999tq, Dvali:2001fw}.
The slingshot effect for such objects will essentially proceed as in the $3+1$-dimensional slingshot systems discussed in the previous sections, with the correction that the tension shall be given by the energy density of the corresponding $D$-brane configuration integrated 
over the extra coordinates $x_{\perp}$. 

In contrast, the transverse slingshot effect is controlled by the $D$-brane dynamics in the $x_{\perp}$ coordinates. The stretching of the string takes place in the direction of the $D$-brane motion. A scenario leading to the production of such stretched strings without considering the slingshot effect was originally discussed in~\cite{Dvali:1999tq}. In the following, we will consider some aspects of a generic transverse slingshot scenario. It should be noted, however, that the slingshot may, in general, consist of a combination of transverse and longitudinal components.

To discuss the essential features of the dynamics of the stretched strings, we mentally uplift the slingshot effect in a single higher dimension.
The role of a ``domain wall" is now played by a $D$-brane that fills the $x_{\parallel}$ coordinates, and is connected to a lower dimensional $D$-brane by a 
string stretched in the $x_{\perp}$ dimensions. 

For the string coupling $g_{\rm s}\sim 1$, the mass of the $D$-brane slingshot is given by the mass of the stretched string, $M_{\rm sl} \sim M_{\rm s}^2L$, where 
$M_{\rm s}$ is the string scale and $L$ is the length. This mass can be expressed in terms of the four-dimensional Planck scale $M_{\rm P}$ as,
\begin{equation}
\label{eq:mass-D-slingshotL}
     M_{\rm sl} \sim  \left(\frac{M_{\rm P}^2}{R_{\perp}}\right)^{\frac{2}{3}} L\,.
\end{equation} 
Taking $L \sim R_{\perp}$ and assuming the compactification radius at its experimental upper bound, $R_{\perp} \sim 10^{-2}$cm, the mass of the slingshot can be as large as
\begin{equation}
\label{eq:mass-D-brane-slingshot}
     M_{\rm sl} \sim  \left(\frac{M_{\rm P}^2}{R_{\perp}}\right)^{\frac{2}{3}} R_{\perp}
     \sim 10^{29}\,\rm GeV\,.
\end{equation} 
Notice that if the slingshot string represents a 
$D_{1+n}$-brane wrapped around additional $n$ extra dimensions of radii $R_{j},~j=1,2,..,,n$, the mass of the slingshot is  
\begin{equation}
 M_{\rm sl} \sim g_{\rm s}^{-1}M_{\rm s}^2 L \,  (2\pi M_{\rm s})^n \prod_{j=1}^n R_{j} \,,
\end{equation}
in accordance with \eqref{Teff}. Taking $n=5$ and using the relation,
$M_{\rm P}^2=g_{\rm s}^{-2}\, M_s^8\,V_6$, where $V_6$ is the total compact volume, we can write the 
upper bound on the mass of the slingshot as 
\begin{equation}
\label{eq:6-dimensions-slingshot-mass}
 M_{\rm sl} \sim g_{\rm s} \, \frac{M_{\rm P}^2}{M_{\rm s}} \,.   
\end{equation} 
For the current experimental lower bound on the quantum gravity scale, $M_{\rm s} \sim\,\rm TeV$~\cite{Arkani-Hamed:1998jmv}, this 
gives $M_{\rm sl} \sim 10^{35}\rm GeV$.

The essential new features are the following. Since the domain walls are now parallel to our dimensions, the question of wall dominance is void. The entire $D$-brane tension accounts for the vacuum energy that drives inflation and later settles to the present vacuum. The total slingshot energy is invested into the creation of inhomogeneities that can source gravitational waves
as well as produce Kaluza-Klein graviton dark matter and PBH dark matter. These inhomogeneities differ from the ones obtained by the four-dimensional mechanisms.

In discussing the peculiarities of $D$-brane slingshots, it is useful to analyze first all three effects separately. Of course, the realistic cosmology is expected to involve all three. At the end of this section, we present an example in $6$ extra dimensions in which all three effects are taken into account.\\ 

\noindent{\bf 1. Gravitational waves in one large extra dimension.}
The characteristic feature is the deformation of the spectrum on the co-moving wavelengths, which at that epoch are $\sim R_\perp$. Let us assume that the slingshot mechanism takes place after the inflationary phase. Then, we get that the redshifted frequency of the current gravitational wave signal is
\begin{align}
    f^{\rm peak}\simeq \frac{T_0}{\sqrt{M_{\rm P}R_\perp}}\,,
\end{align}
where we assumed that the slingshots annihilate when the Universe has a temperature $T_{\rm sl}\sim \sqrt{M_{\rm P}/R_\perp}$. Notice that, for simplicity, we are ignoring factors originating from the number of degrees of freedom. 

The resulting gravitational radiation density parameter can be estimated as
\begin{align}
\label{eq:OmegaGW-Dbrane-Slingshot}
    \Omega^{\rm peak}_{\rm GW}\simeq \epsilon_{\rm GW}\, \frac{N_{\rm sl}M_{\rm sl}H_{\rm sl}}{ M_{\rm P}^2}\, \Omega_{\rm rad}\, ,
\end{align}
where $\epsilon_{\rm GW}$ is the conversion coefficient into gravitational radiation and $N_{\rm sl}$ is the number of slingshots per horizon. The fraction corresponds to the energy budget in the slingshots per horizon compared to the horizon energy at the slingshot time, $M_{\rm P}^2/H_{\rm sl}$. Notice that, if this quantity exceeds one, the Universe is dominated by the slingshot energy at the slingshot time. 

The prefactor $\epsilon$ is fixed by the gravitational coupling. Clearly, this quantity depends on the dimensionality, i.e., whether or not the excited harmonics by the string have a wavelength $\lesssim R_{\perp}$. Since the signal is dominated by harmonics of the largest wavelength, and the system size is initially of length $R_{\perp}$, the $D=4$ coupling,
\begin{equation}
    \epsilon_{\rm GW} \simeq M_{\rm s}^2/M_{\rm pl}^2\,,
\end{equation}
provides a good approximation to the actual peak.

For the analysis of the gravitational radiation emission, several caveats must be kept in mind. The emission of zero-mode gravitons depends sensitively on the orientation of the slingshot within the extra dimension. If the slingshot is exactly perpendicular to the brane, no massless gravitons are produced. However, when the angle between the brane and the slingshot is smaller than $90^\circ$, graviton emission becomes possible. For simplicity, we assume that the slingshot is never perfectly perpendicular to the brane and that the resulting directional contribution to the conversion factor remains of order one. 

Furthermore, the dissipation from extra Kaluza-Klein gravitons might lead to a deviation of the gravitational wave spectrum, which therefore is expected to deviate from the $f^{-1}$ scaling typical for monopoles accelerated by strings. 

In the magnetic monopole slingshot scenario described above, we observed that string-wall junctions can attract each other through Coulomb interactions. This effect limits the duration of the slingshot phase during which gravitational radiation is emitted. In the case of the $D$-brane slingshot, the relevant time scales can instead be constrained by merger effects arising from the gravitational attraction on the brane to which the slingshot is attached. 
As long as the slingshots do not dominate the energy budget of the Universe, we can safely neglect merger effects.\\

\noindent{\bf 2. Kaluza-Klein graviton emission.}
The higher-dimensional slingshot is a natural source for Kaluza-Klein (KK) excitations of the graviton and of other bulk species. As proposed in~\cite{Arkani-Hamed:1998jmv} (see also~\cite{Gonzalo:2022jac} for recent discussions), the Kaluza-Klein gravitons with masses lighter than $m_{\rm max}\sim 100\, {\rm MeV}$ are natural candidates for dark matter, as their lifetime exceeds the current age of the Universe\footnote{The lifetime is estimated as $\tau = \Gamma^{-1} \sim {M_{\rm P}^2}/{m_n^3}$ and is therefore longer than the age of the Universe, $t_0 \sim 10^{10}\,\mathrm{yr}$, for $m_n \lesssim 100\,\mathrm{MeV}$.}. 

As discussed in Section~\ref{sec:gravitational-radiation}, the gravitational radiation spectrum for massless gravitons is expected to scale as $\dd P_{\rm GW}/ \dd\omega \sim \mu^2/(M_{\rm P}^2\, \omega)$, with $\mu$ being the string tension. The above scaling assumes an exactly straight string. However, it is generically expected that the slingshot is curved on length scales comparable to its length. This is sourced by both the interaction between slingshots, c.f. equation~\eqref{eq:slingshot-decay-time}, as well as the relative velocity and impact angle between our $D$-brane and the slingshot. Besides, as previously discussed, this is also a necessary condition for the detectability of gravitational waves.  
This implies that the slingshot radiates with a frequency dependence $\propto \omega^{-1}$ up to a critical frequency $\omega_* \simeq 1/L_* = n_* / R_{\perp}$, where $n_*$ denotes the critical harmonic associated with the length scale $L_*$ characterizing the bending of the string. At higher frequency, the spectrum is further suppressed~\cite{Martin:1996cp} and is therefore ignored in the following discussion.

Due to their mass, KK gravitons can only be emitted at frequencies $\omega > m_n$, where $m_n = n/R_{\perp}$ denotes the mass of the $n$th graviton mode.
Since the emission into a given KK mode is of the same order as the power emitted into the graviton zero mode for $\omega > m_n$, we have for a single slingshot,
\begin{align}
\label{eq:en}
E_n \sim  \frac{\mu^2}{M_{\rm P}^2}\, R_\perp \sum_{k=n}^{n_*} \frac{1}{k}\,,
\end{align}
where $n_*$ is set by the slingshot curvature.

Notice that this energy includes both the rest mass and the kinetic energy of the gravitons. 
Only the rest mass contribution from the lightest KK gravitons 
\begin{align}
    E_n\sim \frac{\mu^2}{M_{\rm P}^2} \,  \frac{1}{m_n}\, , 
\end{align}
sizably accounts for the dark matter abundance. The corresponding energy density parameter $\Omega_n$ associated with the $n$th KK graviton mode is given by
\begin{align}
    \Omega_n \sim \frac{M_{\rm P}^{\frac{1}{2}}}{m_n\, R_\perp^{\frac{3}{2}}\,T_0}\, \Omega_{\rm GW}^{\rm peak}\, ,
\end{align}
where $\Omega_{\rm GW}^{\rm peak}$ corresponds to the density parameter of the massless zero mode given in equation~\eqref{eq:OmegaGW-Dbrane-Slingshot}.

Summing over all graviton modes that can contribute to the present-day dark matter abundance leads to the total density parameter
\begin{align}
\label{eq:OmegaKK}
\Omega_{\rm KK} \sim \sqrt{\frac{M_{\rm P}}{R_\perp}}\, \frac{1}{T_0}\ln \left({\rm min}[m_{\rm max} R_\perp,n_*]\right)\, \Omega_{\rm GW}^{\rm peak}\, ,
\end{align}
This result provides a useful approximate bound on the gravitational radiation density parameter, since the KK graviton dark matter contribution should not exceed $\Omega_{\rm KK} \sim \mathcal{O}(1)$.

Notice again that only KK gravitons with masses lighter than $m_{\rm max}\sim 100\, \rm MeV$ can contribute to the present-day dark matter abundance. Therefore, KK gravitons can be produced only if the slingshot takes place in an extra dimension of size $R_\perp\gtrsim 10^{-13}\, \rm cm$. Furthermore, equation~\eqref{eq:OmegaKK} is also valid in the case of a perfectly straight string, corresponding to $n_* \gg 1$. 

One caveat must be taken into account when considering the emission of KK gravitons. In principle, the KK graviton emission can influence the slingshot dynamics. Consequently, the formulas given above are applicable if only a small fraction of the total slingshot mass is converted into massless or massive gravitational radiation. In the regimes we are considering - for which $n_* \sim \mathcal{O}({\rm few})$ - the backreaction is negligible.\\

\noindent{\bf 3. PBH formation.} As discussed in~\cite{Dvali:2021byy}, the slingshot mechanism can lead to the formation of PBHs due to the re-acceleration of the $D$-branes by the string tension. Indeed, consider a string that connects ``our" $D$-brane, oriented along $x_{\parallel}$, with a lower dimensional one, displaced along $x_{\perp}$ (see Fig.~\ref{fig:magnetic-energy-density}). The string tension pulls the point-like $D$-brane towards our brane, converting the energy of string tension into the kinetic energy of the accelerated $D$-brane. Upon collision with our $D$-brane, a black hole is formed. For a maximal initial length of the slingshot, the mass of the resulting PBH is given by~\eqref{eq:mass-D-brane-slingshot}, with the $5$-dimensional gravitational radius 
\begin{equation}
\label{r5}
  R_{5} \sim  \left( M_{\rm sl} \frac{R_{\perp}}{M_{\rm P}^2} \right )^{\frac{1}{2}} \sim 
   \frac{R_{\perp}^{\frac{2}{3}}}{M_{\rm P}^{\frac{1}{3}}} \sim \sqrt{\frac{R_\perp}{M_{\rm s}}}\;. 
\end{equation}
This relation implies that the $5$-dimensional gravitational radius of the object, $R_5$, is much larger than the string scale, $M_{\rm s}^{-1}$. The latter scale sets the thickness of the $D$-branes as well as of the string connecting them. Correspondingly, in a head-on collision, the $D$-brane slingshot is expected to produce a black hole with high probability, since the center of mass energy gets localized within its $5$-dimensional gravitational radius. 

Remarkably, as it is clear from equation~\eqref{eq:mass-D-brane-slingshot}, for the experimentally-viable values of the compactification radius $R_{\perp} \sim 10^{-2} $cm~\cite{Adelberger:2009zz, Tan:2016vwu}, this could result in PBHs as heavy as $10^5\,{\rm g}$. 

\interfootnotelinepenalty=10000
These black holes would have an evaporation timescale of order $10^{7}\,{\rm s}$ according to the naive extrapolation of the Hawking rate~\cite{Hawking:1974rv} throughout the entirety of the lifetime. However, as realized recently, the lifetime is expected to be substantially prolonged due to the \textit{memory burden} effect, which takes into account the backreaction from the information load carried by the black hole~\cite{Dvali:2018xpy, Dvali:2020wft, Alexandre:2024nuo, Dvali:2024hsb,Dvali:2025ktz} (for recent updates, see~\cite{Dvali:2025sog} and references therein). With this effect taken into account, the lifetime of the above black hole would be prolonged to at least $10^{20}$ times the present age of the Universe.\footnote{Notice that even in the semiclassical regime the higher-dimensional black holes, with radius $R_5 \gg R_{\perp}$, live longer than the $4$-dimensional ones of the same mass. From the $4$-dimensional perspective, this can be viewed as the effect of the species hair~\cite{Dvali:2008fd,Dvali:2008rm}. The role of the species is played by Kaluza-Klein gravitons as well as string excitations~\cite{Dvali:2007hz, Dvali:2007wp}. This fact was used for obtaining dark matter from higher-dimensional PBHs in~\cite{Anchordoqui:2022txe, Anchordoqui:2025opy} (see also~\cite{Ettengruber:2025kzw}). However, this possibility does not work for the 
above-discussed slingshot black holes with masses $10^5\,{\rm g}$. These will experience the memory burden effect latest after $10^{7}\,{\rm s}$~\cite{Dvali:2025ktz} and will likely get stabilized by it.}
\interfootnotelinepenalty=0
   
In this light, the microscopic PBHs produced by the slingshot mechanism fall in the range of masses relevant for astrophysical high-energy $\gamma$-ray and neutrino observations~\cite{Fermi-LAT:2014ryh,LHAASO:2023gne,IceCube:2018fhm,IceCube:2020wum,KM3NeT:2025npi}.

Assuming that a fraction $\epsilon_{\rm PBH}$ of the slingshots forms a black hole, the corresponding energy density parameter for PBHs is given by~\cite{Dvali:2021byy}
\begin{align}
\label{eq:PBH-density-parameter-michael-paper}
    \Omega_{\rm PBH}\sim \epsilon_{\rm PBH}\,N_{\rm sl}\,\frac{\mu^{\frac{3}{2}}}{M_{\rm BH}^{\frac{1}{2}}\, T_0\, M_{\rm P}^{\frac{3}{2}}}\, \Omega_{\rm rad}\,,
\end{align}
where $M_{\rm BH}=M_{\rm sl}=\mu R_\perp$ is the mass of the black holes. For one (six) extra dimension(s), $M_{\rm BH}\lesssim 10^{5}\,\rm g$ ($M_{\rm BH}\lesssim 10^{11}\,\rm g$) according to the discussion around equation~\eqref{eq:mass-D-brane-slingshot} (equation~\eqref{eq:6-dimensions-slingshot-mass}). The conversion factor $\epsilon_{\rm PBH}$ accounts for the fact that not all the slingshots might form a PBH due to dynamical reasons, such as bending or interaction with other D-branes. Notice that most of the slingshots could collapse without turning into a black hole. This type of disappearance can take place either due to annihilation or erasure~\cite{Dvali:1997sa}. 

Notice that the PBH abundance in equation~\eqref{eq:PBH-density-parameter-michael-paper} is the same as the one discussed in~\cite{Dvali:2021byy}. In that work, PBHs were formed via the collapse of confined quarks connected by flux tubes of length comparable to the horizon size. In the present work, we assumed that the size of the slingshot is given by the horizon scale at the time of the PBH formation as well. Relaxing this hypothesis leads to a different parametric behavior.\\

\noindent{\bf Example: Slingshot in six extra dimensions.}
Now, let us consider six extra dimensions, as in standard string theory. We assume that one of the extra dimensions is compactified at the scale $R_\perp$, while the remaining five are compactified at scales $R_j \ll R_\perp$. Instead of a fundamental string attached to ``our" $4$-dimensional brane, we consider a $D_{1+5}$-brane that is wrapped around the five small compact dimensions. Effectively, this configuration behaves as a string in the large extra dimension, with an effective string tension given by
\begin{align}
\mu \simeq \frac{M_{\rm P}^2}{R_\perp M_{\rm s}}\, .
\end{align}
The gravitational radiation conversion factor is
\begin{align}
    \epsilon^{6d}_{\rm GW}\simeq \frac{\mu}{M_{\rm P }^2}\simeq \frac{1}{M_{\rm s} R_\perp}\, ,
\end{align}
leading to the gravitational wave density parameter
\begin{align}
\label{eq:OmegaGW6ExtraDimensions}
    \Omega_{\rm GW}^{\rm peak}\simeq \frac{N_{\rm sl}}{M_{\rm s}^2 R_\perp^2}\, \Omega_{\rm rad}\simeq 10^{-13}\, N_{\rm sl}\left(\frac{\rm TeV}{M_{\rm s}}\right)^2 \left(\frac{f}{10\, {\rm Hz}}\right)^4\, .
\end{align}
The condition to avoid energy domination by the slingshot is given by
\begin{align}
\label{eq:maximal-slingshot-number-Dbrane-slingshot}
    N_{\rm sl}\lesssim \frac{M_{\rm P}}{M_{\rm sl}H_{\rm sl}}\simeq M_{\rm s}R_\perp \simeq 10^4 \, \frac{M_{\rm s}}{\rm TeV}\left(\frac{10\, {\rm Hz}}{f}\right)^2\, .
\end{align}

For one slingshot per horizon, $M_{\rm s}\sim \rm TeV$, and $f \sim 10\, {\rm Hz}$ ($R_\perp\sim 10^{-13}\, {\rm cm}$), the signal is in the right range of the future generation of gravitational wave experiments like the Einstein Telescope~\cite{Sathyaprakash:2012jk} and Cosmic Explorer~\cite{LIGOScientific:2016wof}.
Increasing the number of slingshots per horizon to the maximally permitted value of $N_{\rm sl} \sim 10^4$ allows for a signal that may even fall within the sensitivity range of current detectors such as LVK~\cite{aLIGO:2020wna,VIRGO:2014yos,KAGRA:2018plz}.

Notice that equation~\eqref{eq:OmegaGW6ExtraDimensions} can be only applied for frequencies $f \gtrsim 10\, \rm Hz$ ($R_\perp \lesssim 10^{-13}\, \rm cm$), because for lower frequencies the system enters the regime in which KK gravitons are produced during the slingshot. 

We can use equation~\eqref{eq:OmegaKK} to estimate the KK energy density parameter for $R_\perp \gtrsim 10^{-13}\,\rm cm$. For the setup discussed in this example, it is then given by
\begin{align}
    \Omega_{\rm KK}\sim 10^{-1}\,N_{\rm sl}\, \left(\frac{10^{-9}\,\rm cm}{R_\perp}\right)^{\frac{5}{2}}\left(\frac{\rm TeV}{M_{\rm s}}\right)^{2}\, ,
\end{align}
where the logarithm in equation~\eqref{eq:OmegaKK} was dropped.
Therefore, the emission of KK gravitons can lead to density parameters comparable to the present-day dark matter energy density.

Notice that, for the values we are considering, the energy emitted into gravitons is less than one percent of the slingshot energy budget. Therefore, the backreaction on the slingshot from graviton emission is negligible.

The PBH density parameter for $\mu \simeq M_{\rm P}^2/(R_{\perp} M_{\rm s})$ is 
\begin{equation}
    \Omega_{\rm PBH} \sim \epsilon_{\rm PBH}\, N_{\rm sl}\,\frac{M_{\rm P}^{1/2}}{M_{\rm s}\, R_{\perp}^{3/2}\, T_0 }\,\Omega_{\rm rad}\,.
\end{equation}
For $R_{\perp} = 10^{-13}\,\rm cm$ and $M_{\rm s}\sim \rm TeV$, only a fraction of $\epsilon_{\rm PBH}\, N_{\rm sl}^{-1}\sim 10^{-14}\, N_{\rm sl}^{-1}$ of the annihilating slingshot energy budget needs to be converted into black holes to account for the observed dark matter abundance in our Universe. The resulting mass for this choice of parameters is $M_{\rm BH}\sim 10^{11}\,\rm g$. 

Notice that for a larger $M_{\rm s}\gtrsim \rm TeV$ the mass is proportionally lowered. According to the semiclassical picture, such PBHs are too light to constitute the dark matter. However, they can be stabilized by the memory burden effect~\cite{Dvali:2018xpy, Dvali:2020wft, Alexandre:2024nuo, Dvali:2024hsb,Dvali:2025ktz} potentially bearing spectacular astrophysical signatures in the form of cosmic rays, see e.g.,~\cite{Zantedeschi:2024ram,Dvali:2024hsb,Dondarini:2025ktz}. The emission temperature of these objects is $M_{\rm s}$, and therefore they emit not only into Standard Model degrees of freedom, but also into KK modes, potentially altering the emitted energy budget available for detection. A complementary study accounting for such effects is currently in progress~\cite{Zantedeschi2026toappear}.

\section{Discussion and Outlook}
\label{sec:discussion-and-outlook}

The slingshot effect~\cite{Bachmaier:2023wzz} takes place when a flux-carrying source crosses a boundary between domains with a confined and unconfined flux. In the confining domain, the flux forms a tube pulling the source towards the domain in which the flux can freely spread. The phenomenon is rather general and can take place for both magnetic and electric fluxes. The role of the flux tube can be played by cosmic strings, QCD-like strings, as well as by fundamental strings and $D$-strings of string theory. 

In this paper, we have studied the generalizations and various implications of the slingshot effect, in particular, taking into account interactions between multiple slingshots. 

The slingshot effect can take place 
for sources and fluxes of different co-dimensionalities. For example, a cosmic string in $3+1$ dimensions can be ``slingshotted" by a domain wall~\cite{Bachmaier:2023wzz}. 
Likewise, $D$-branes of various dimensionalities can be subjected to a slingshot effect by fundamental strings and/or $D$-branes of lower co-dimensionalities. 

In the present paper, we studied the prototype model of the slingshot for magnetic monopoles. However, the study equally extends to a dual scenario involving the case of the ``electric" confinement. 
Here, we consider the passage of a heavy quark through the domain wall separating the $U(1)$ Coulomb and the confining QCD vacua. Such a setup was originally discussed in 
\cite{Dvali:2002fi, Dvali:2007nm} using the mechanism of \cite{Dvali:1996xe} for creating the coexisting confining and Coulomb phases of an $SU(N)$ gauge theory separated by a domain wall. 
 
A heavy quark entering the confining phase from the Coulomb phase gets connected to the wall by a QCD string. The flux transported by this string diffuses into the Coulomb region, effectively matching the electric charge of the initial quark.
In the absence of the light quarks, the string  
shall not break (hadronize) easily. More recent discussions of such setups in the cosmological context can be found in~\cite{Baldes:2020kam, Gouttenoire:2023roe}. 

Notice, as shown in~\cite{Dvali:2007nm}, when a Coulomb phase layer of finite width is ``sandwiched" between the two parallel confining phases, the $U(1)$ effectively becomes confining inside the layer at exponentially large distances due to the effect of virtual monopoles tunneling across the Coulomb layer.

\begin{figure}
    \includegraphics[width=1\linewidth]{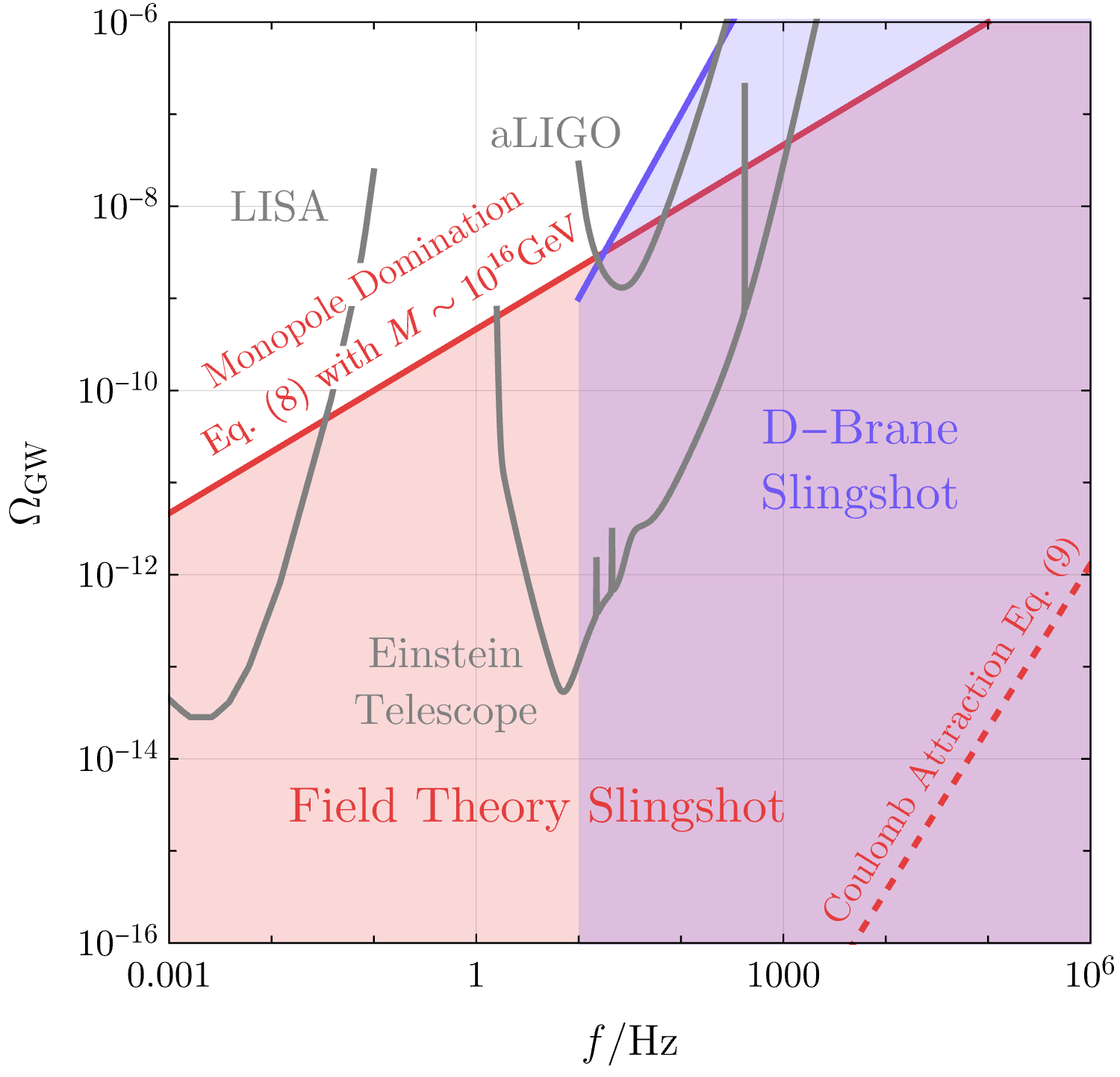}
    \caption{\justifying The red area indicates the region in which the field-theoretic slingshot (monopole slingshot) can potentially generate a gravitational radiation signal. The solid red line corresponds to the bound arising from monopole domination for a monopole mass of $M \sim 10^{16}\,\mathrm{GeV}$ (see equation~\eqref{eq:monopole-domination}). For parameters below the dashed red line, the Coulomb interaction is relevant on slingshot timescales (see equation~\eqref{eq:condition-collapse-faster-than-Coulomb-attraction}).
    The blue region shows the regime in which the wrapped $D$-brane slingshot can produce a gravitational radiation signal. The blue line illustrates the bound imposed by the requirement that the slingshots do not dominate the energy budget of the Universe (see equation~\eqref{eq:maximal-slingshot-number-Dbrane-slingshot}). The sensitivity curves for LISA~\cite{LISA:2017pwj}, aLIGO~\cite{aLIGO:2020wna}, and the Einstein Telescope~\cite{Sathyaprakash:2012jk} were obtained using the open-access code provided in~\cite{Mingarelli:2019mvk}. A Hubble constant of $68\,\mathrm{km/s/Mpc}$ was adopted.}
    \label{fig:OmegaGW}
\end{figure}

We expect the slingshot effect to take place in a broad class of theories, those in which the early Universe is characterized by phase transitions where both Higgsed (confining) and non-Higgsed (unconfining) vacua can coexist at the same time. As we discussed in this work, once confined charges are present in the theory, the dynamics of first-order phase transitions change significantly, leading to relevant modifications of today's expected gravitational radiation from the early Universe. Furthermore, the gravitational wave signal from the slingshot itself can provide a unique signature. In Figure~\ref{fig:OmegaGW} we show the regions in which the monopole slingshot may emit gravitational radiation (red area). We observe that the resulting signal falls within the sensitivity range of current and future gravitational-wave experiments such as LISA~\cite{LISA:2017pwj}, aLIGO~\cite{aLIGO:2020wna}, and the Einstein Telescope~\cite{Sathyaprakash:2012jk}.
 
We wish to comment that the phase portrait in the epoch relevant for the slingshot effect can be much richer, consequently changing the predictions for gravitational waves and other observables. In particular, the gauge coupling in the early Universe is expected to be different from its today's value both due to changes of the field masses as well as their expectation values~\cite{Dvali:1995ce}. This can have multiple important effects. If the coupling is pushed towards strong values, it can substantially increase the confinement scale for QCD-like strings, thereby enhancing the slingshot effect for confined heavy quarks. Similarly, the push towards weak coupling will increase the tension of magnetic strings, as well as the mass of magnetic monopoles. 

Another effect that could have interesting consequences is the ``melting" of the domain walls~\cite{Vilenkin:1981zs} (for more recent analysis, see~\cite{Babichev:2021uvl,Ramazanov:2021eya,Babichev:2023pbf}). If the melting takes place during the slingshot effect, this can dwarf the contribution to gravitational waves from domain walls relative to confined monopoles (or quarks). More detailed discussion of such effects is left for future work.  

Finally, we have discussed that one natural arena for the implementation of the slingshot effect is inflationary cosmology in string theory, which is generically based on $D$-branes~\cite{Dvali:1998pa, Dvali:1999tq, Dvali:2001fw}. 
In particular, in $D$-brane cosmology, the production of the long strings that are attached to the $D$-branes is rather generic~\cite{Dvali:1999tq}. 

Our analysis shows that the relative motion of $D$-branes, 
which is accompanied by stretched strings that connect them, results in the slingshot effect, generically accompanied by the emission of gravitational waves and the production of KK gravitons in the form of dark matter. In Figure~\ref{fig:OmegaGW}, the regime in which the $D$-brane slingshot can emit gravitational radiation is illustrated (blue area).
In addition, the subsequent pull-over of the $D$-branes by the string tension results in the formation of PBHs via the mechanism of~\cite{Dvali:2021byy}. The characteristic mass range is in the sub-asteroid domain. In the light of stabilization by the memory burden effect~\cite{Dvali:2018xpy,Dvali:2020wft,Alexandre:2024nuo,Dvali:2024hsb,Dvali:2025ktz}, such black holes can be viable candidates for dark matter. Since light burdened PBHs emit high energy particles~\cite{Dvali:2020wft, Alexandre:2024nuo,Thoss:2024hsr,Dvali:2024hsb,Dvali:2025ktz}, they represent interest for high energy cosmic ray observatories~\cite{Chianese:2024rsn,Zantedeschi:2024ram,Boccia:2025hpm,Dvali:2025ktz,Tan:2025vxp,Dondarini:2025ktz}.\\

\section*{Acknowledgments}
This work was supported in part by the Humboldt Foundation
under Humboldt Professorship Award, by the European
Research Council Gravities Horizon Grant AO number:
850 173-6, by the Deutsche Forschungsgemeinschaft
(DFG, German Research Foundation) under Germany's
Excellence Strategy - EXC-2111 - 390814868, and
Germany's Excellence Strategy under Excellence Cluster
Origins.

J.S.V.B. acknowledges support from the Spanish Ministry of Science and Innovation (MICINN) through the Spanish State Research Agency under the R\&D\&I project PID2023-146686NB-C31 funded by MICIU/AEI/10.13039/501100011033/ and by ERDF/EU, and under Severo Ochoa Centres of Excellence Programme 2025-2029 (CEX2024001442-S). The CERCA program of the Generalitat de Catalunya partially funds IFAE. MICIIN supported this study with funding from the European Union NextGenerationEU (PRTR-C17.I1) and by the Generalitat de Catalunya. This work was supported by the Juan de la Cierva program JDC2024-055046-I, funded by MICIU/AEI/10.13039/501100011033 and the ESF+.

Disclaimer: Funded by the European Union. Views and opinions expressed are, however, those of the authors only and do not necessarily reflect those of the European Union or European Research Council. Neither the European Union nor the granting authority can be held responsible for them.

\setlength{\bibsep}{4pt}
\bibliography{references}

\end{document}